\documentclass[12pt]{article}
\usepackage{a4wide,amssymb,cite}
\pdfoutput=1

\usepackage{a4wide,amssymb,graphicx}
\usepackage{epsfig}
\usepackage[usenames,dvipsnames]{color}
\usepackage{slashed}
\usepackage{amssymb,cite,graphicx}
\usepackage{slashed}
\usepackage{amsmath,bm,bbm}
\usepackage{amsfonts}
\usepackage[titletoc,title]{appendix}
\usepackage[small]{caption}
\usepackage[margin=1in]{geometry}
\usepackage[multiple]{footmisc}
\usepackage{mathtools}
\usepackage{slashed}
\usepackage[nottoc]{tocbibind}
\usepackage{xcolor}

\newcommand{\be}{\begin{equation}}
\newcommand{\ee}{\end{equation}}
\newcommand{\bea}{\begin{eqnarray}}
\newcommand{\eea}{\end{eqnarray}}

\def\circa#1{\,\raise.3ex\hbox{$#1$\kern-.75em\lower1ex\hbox{$\sim$}}\,}

\begin{document}

\begin{titlepage}
%
%

%\rightline{CERN-TH-2020-xxx}

%

\begin{centering}
\vspace{1cm}
{\Large {\bf Peccei-Quinn Inflation at the Pole \vspace{0.15cm} \\  and Axion Kinetic Misalignment}} \\

\vspace{1.5cm}

\begin{centering}
{\bf Hyun Min Lee$^{1,\dagger}$, Adriana G. Menkara$^{1,2,\sharp}$, \vspace{0.15cm} \\  Myeong-Jung Seong$^{1,\ddagger}$, and Jun-Ho Song$^{1,\star}$ }
\end{centering}
\\
\vspace{.5cm}

{\it $^1$Department of Physics, Chung-Ang University, Seoul 06974, Korea.}
\\
{\it $^2$CERN, Theory department, 1211 Geneva 23, Switzerland.}

\vspace{.5cm}

%\today

\end{centering}
\vspace{2cm}

\begin{abstract}
\noindent
We propose a minimal extension of the Standard Model with the Peccei-Quinn (PQ) scalar field and explain the relic density of the QCD axion through the kinetic misalignment with a relatively small axion decay constant. To this purpose, we consider a slow-roll inflation from the radial component of the PQ field with the PQ conserving potential near the pole of its kinetic term and investigate the post-inflationary dynamics of the PQ field for reheating. The angular mode of the PQ field, identified with the QCD axion, receives a nonzero velocity during inflation due to the PQ violating potential, evolving with an approximately conserved Noether PQ charge. We determine the reheating temperature from the perturbative decays and scattering processes of the inflaton and obtain dark radiation from the axions produced from the inflaton scattering at a testable level in the future Cosmic Microwave Background experiments. We show the correlation between the reheating temperature, the initial velocity of the axion and the axion decay constant, realizing the axion kinetic misalignment for the correct relic density.

\end{abstract}

\vspace{2.5cm}

\begin{flushleft} 
$^\dagger$Email: hminlee@cau.ac.kr \\
${}^{\sharp}$Email: amenkara@cau.ac.kr  \\
$^\ddagger$Email: tjdaudwnd@gmail.com  \\
$^\star$Email: thdwnsgh1003@naver.com
\end{flushleft}

\end{titlepage}

\section{Introduction}

%motivation
The QCD axion has been proposed to solve the strong CP problem as a pseudo-Goldstone boson appearing after an anomalous U(1) Peccei-Quinn (PQ) symmetry is broken spontaneously \cite{PQ,KSVZ,DFSZ}. The QCD axion is also a good candidate for cold dark matter whose abundance is generated by the misalignment mechanism \cite{misalign}.  If the axion was once in thermal equilibrium due to a relatively high temperature of the universe, it could be not only a dominant component for dark matter, but also become a component for dark radiation \cite{thermalaxions1,thermalaxions2,DR}, which is testable in the future experiments for Cosmic Microwave Background (CMB) anisotropies.

%content
Recently, the general initial conditions with a nonzero velocity for the QCD axion or axion-like particles have been considered, under the name of the axion kinetic misalignment mechanism \cite{Co:2019jts}.  At the onset of the axion oscillation, the kinetic energy for the axion is comparable to the potential energy for the axion, so the axion abundance depends on the axion velocity initially set by the Noether PQ charge in the early universe.  The PQ symmetry is broken explicitly by quantum gravity effects, so it naturally induces a nonzero velocity for the axion. However, since the overall magnitude of the PQ violating potential is set by the modulus of the PQ field, we need to take into account the dynamics of the PQ field in the early universe and make sure that the consistent picture with the axion kinetic misalignment at a later stage emerges. We remark that the axion dynamics with a nonzero initial velocity has drawn much attention for axiogenesis \cite{axiogenesis,axio1}, gravitational waves \cite{GW}, axion fragmentation \cite{axionfrag}, etc.

%results
In this article, we consider the extension of the SM with a PQ complex scalar field in the presence of the U(1) PQ symmetry. The angular component of the PQ field is identified as the QCD axion and the radial component of the PQ field is regarded as the saxion. The QCD anomalies are generated due to an extra vector-like quark as in the KSVZ axion model \cite{KSVZ}, providing the axion-gluon coupling necessary for the dynamical solution to the strong CP problem. In this model, we pursue the origin of the initial condition for the axion misalignment in the context of the PQ inflation at the pole \cite{alpatt,Clery:2023ptm}. Then, in contrast to the previous studies on the PQ inflation with a small non-minimal coupling \cite{PQ0,PQ1,PQ2}, the saxion couples conformally to gravity, providing a possibility of realizing the PQ inflaton near the pole of the kinetic term in the Einstein frame, without relying on trans-Planckian values of the PQ field for inflation. The dynamics of the saxion is dominated by the PQ conserving potential.  On the other hand, we show that the PQ violating potential sets the initial velocity for the axion during inflation, and an approximately conserved Noether PQ charge is achievable at the end of reheating. 

We compute a sufficiently large reheating temperature of the universe from the decays and scattering processes of the saxion and identify the dark radiation component from the axions produced by the saxion scattering. We show the condition for the successful axion kinetic misalignment  in the parameter space for the reheating temperature, the initial velocity of the axion and the axion decay constant (or the axion mass).

%paper
The paper is organized as follows. 
We first present the model setup for the PQ inflation and the interactions of the PQ field to the extra vector-like quark, the SM Higgs, and possibly to the right-handed neutrinos. In this model, we discuss the Einstein-frame potential, the bounds on the PQ violating potential from the axion quality and the axion-photon coupling. Next we discuss the inflationary dynamics depending on whether inflation is driven by the PQ conserving or violating potentials, and describe the post-inflationary dynamics of the inflaton and the axion from the classical equations of motion. We continue to consider the reheating period in the presence of the decays and scattering processes of the inflaton and determine the reheating temperature as well as the dark radiation from the axions accordingly. From the initial axion velocity given at the end of inflation, we show the evolution of the PQ Noether charge at reheating and find the parameter space where the axion kinetic misalignment is a dominant production mechanism for axion dark matter.
Finally, conclusions are drawn.

\section{The setup}

We introduce a complex scalar field $\Phi$, which is charged under the U(1) PQ symmetry. We assume that the PQ field is conformally coupled to gravity \cite{Clery:2023ptm}, but the conformal symmetry is broken by the mass term for the PQ field and the higher order interactions in the PQ potential. Then, we consider the Jordan-frame Lagrangian for the PQ inflation as
\bea
\frac{{\cal L}_J}{\sqrt{-g_J}} = -\frac{1}{2}M^2_P\, \Omega(\Phi) R(g_J) + |\partial_\mu \Phi|^2 -\Omega^2(\Phi) V_E(\Phi) \label{LJ}
\eea
where the non-minimal coupling function and the PQ potential are taken to
\bea
\Omega(\Phi) &=& 1-\frac{1}{3M^2_P}|\Phi|^2, \\
V_E(\Phi)&=& V'_0+\frac{\beta_m}{M^{2m-4}_P} |\Phi|^{2m} -m^2_\Phi |\Phi|^2+  \bigg(\sum_{k=0}^{[n/2]}\frac{c_k}{2M^{n-4}_P} \, |\Phi|^{2k}\Phi^{n-2k} +{\rm h.c.}\bigg).  \label{potential}
\eea
Here, $V_0$ is the cosmological constant, $\beta_m$ is a PQ conserving dimensionless parameter, $c_k$ are dimensionless parameters, parametrizing the explicit breaking of the PQ symmetry at the Planck scale \cite{axionquality} in the general form of order $n>4$, and $[n/2]$ is the largest integer smaller than $n/2$.  The PQ violating terms of order higher than $n$ can be added, but we assume that they are suppressed for the inflaton or axion potentials. 

For $m=2$, we can recast the PQ conserving part of the potential in eq.~(\ref{potential}) into
\bea
V_{\rm PQ} = V_0+ \lambda_\Phi\bigg(|\Phi|^2-\frac{f^2_a}{2}\bigg)^2
\eea
where the parameters are redefined as $\beta_m=\lambda_\Phi$, $m^2_\phi=\lambda_\Phi f^2_a$ and $V'_0=V_0 +\frac{1}{4}\lambda_\Phi f^4_a$.

In the KSVZ models \cite{KSVZ}, it is necessary to introduce the Yukawa coupling of the PQ field to a vector-like extra quark $Q$ for the QCD anomalies, as follows,
\bea
{\cal L}_{Q,{\rm int}} = - y_Q \Phi {\bar Q}_R Q_L +{\rm h.c.}
\eea
where $Q_L, Q_R$ carries $+1$ and $-1$ and $\Phi$ carries $-2$ under the $U(1)_{\rm PQ}$ symmetry.  
We can also introduce a Higgs-portal quartic coupling between the SM Higgs and the PQ field, as follows,
\bea
\Delta V_E = \lambda_{H\Phi} |\Phi|^2 |H|^2. 
\eea
As a consequence, the Higgs-portal quartic coupling and the Yukawa couplings for the PQ field are crucial for the decays or scattering of the saxion inflaton after inflation and determining the reheating temperature, as will be discussed in the later section. 

We also comment on the possibility that right-handed neutrinos, $N_i (i=1,2,3)$, carry PQ charge $+1$, and they receive masses after the PQ symmetry is broken.
In this case, we can also introduce extra Yukawa couplings of the PQ field for the Majorana masses of the right-handed neutrinos as
\bea
{\cal L}_{N,{\rm int}} = -\frac{1}{2}\lambda_N \Phi \overline {N^c_L} N_R +{\rm h.c.}.
\eea 
Then, in order to generate neutrino masses by seesaw mechanism, we need to include the Yukawa couplings between the right-handed neutrinos and the SM lepton doublets by ${\cal L}_{\rm seesaw}=-y_N {\tilde H} {\bar l}_L N_R+{\rm h.c.}$, so we need to assign a PQ charge $+1$ to the lepton doublets. But, we can assign the vector-like PQ charges for the SM charged leptons, so they do not contribute to the electromagnetic anomalies for the PQ symmetry. If the right-handed neutrinos are neutral under the PQ symmetry, the right-handed neutrinos couple to the PQ field by non-renormalizable interactions or through the graviton exchanges.

\subsection{Einstein frame Lagrangian}

After making a Weyl transformation of the metric by $g_{J,\mu\nu}=g_{E,\mu\nu}/\Omega$ with $\Omega=1-\frac{1}{3M^2_P}|\Phi|^2$, we obtain the Einstein-frame Lagrangian as follows,
\bea
\frac{{\cal L}_E}{\sqrt{-g_E}} &=&-\frac{1}{2} M^2_P R(g_E) + \frac{|\partial_\mu \Phi|^2}{\big(1-\frac{1}{3M^2_P}|\Phi|^2\big)^2} \nonumber \\
&&-\frac{1}{3M^2_P\big(1-\frac{1}{3M^2_P}|\Phi|^2\big)^2}\bigg(|\Phi|^2 |\partial_\mu \Phi|^2-\frac{1}{4}\partial_\mu |\Phi|^2 \partial^\mu |\Phi|^2\bigg) -V_E(\Phi). \label{LE}
\eea

We take the PQ field in the polar coordinate representation,
\bea
\Phi=\frac{1}{\sqrt{2}}\,\rho\,e^{i\theta}
\eea
where $\rho,\theta$ are radial and angular modes of the PQ field, respectively.
Then, choosing $m=2$ for the PQ conserving potential in eq.~(\ref{potential}), we get the Einstein-frame Lagrangian for the radial  mode $\rho$ and the angular mode $\theta$ as
\bea
\frac{{\cal L}_E}{\sqrt{-g_E}} =-\frac{1}{2} M^2_P R +\frac{1}{2}\,\frac{(\partial_\mu \rho)^2}{\big(1-\frac{1}{6M^2_P}\rho^2\big)^2}+\frac{1}{2}\frac{\rho^2 (\partial_\mu\theta)^2}{\big(1-\frac{1}{6M^2_P}\rho^2\big)} - V_E(\rho,\theta) \label{Linf}
\eea
where
\bea
V_E(\rho,\theta)=V_0 + \frac{1}{4} \lambda_\Phi (\rho^2-f^2_a)^2 +\frac{\rho^n}{2^{n/2}M^{n-4}_P} \sum_{k=0}^{[n/2]} |c_k|\, \cos\Big((n-2k)\theta+A_k\Big),
\eea
with $c_k=|c_k| e^{iA_k}$.
Making the  kinetic term for the radial mode canonically normalized by
\bea
\rho=\sqrt{6}M_P \tanh\Big(\frac{\phi}{\sqrt{6}M_P}\Big), \label{can}
\eea
we rewrite the Einstein-frame Lagrangian in eq.~(\ref{Linf}) as
\bea
\frac{{\cal L}_E}{\sqrt{-g_E}} =-\frac{1}{2} M^2_P R + \frac{1}{2}(\partial_\mu\phi)^2 +3M^2_P\sinh^2\Big(\frac{\phi}{\sqrt{6} M_P}\Big)\,(\partial_\mu\theta)^2- V_E(\phi,\theta), \label{Elag}
\eea
where
\bea
V_E(\phi,\theta)&=& V_{\rm PQ}(\phi) +V_{\rm PQV}(\rho,\theta),  \label{Einpot}
\eea
with
\bea
V_{\rm PQ}(\phi) &=&V_0 +  \frac{1}{4} \lambda_\Phi \Big(6M^2_P  \tanh^2\Big(\frac{\phi}{\sqrt{6}M_P}\Big)-f^2_a\Big)^2, \\
V_{\rm PQV}(\rho,\theta) &=& 3^{n/2}M^4_P \tanh^n\Big(\frac{\phi}{\sqrt{6}M_P}\Big) \sum_{k=0}^{[n/2]} |c_k| \cos\Big((n-2k)\theta+A_k\Big).
\eea
We can also take the VEV of the PQ field in the true vacuum to satisfy $6M^2_P  \tanh^2\big(\frac{\langle\phi\rangle}{\sqrt{6}M_P}\big)\simeq f^2_a$, so $\langle\rho\rangle\sim\langle\phi\rangle\sim f_a$ for $f_a\ll M_P$. After the angular mode gets a VEV $\langle\theta\rangle$, we can get an observed tiny cosmological constant by tuning  the bare cosmological constant $V_0$ to cancel the vacuum energy coming from the PQ breaking terms, as follows,
\bea
V_0\simeq - M^4_P\bigg(\frac{f_a}{\sqrt{2}M_P}\bigg)^n \,\sum_{k=0}^{[n/2]} |c_k| \cos\Big((n-2k)\langle\theta\rangle+A_k\Big). 
\eea 
Moreover, the kinetic term for the axion in the true vacuum becomes $3M^2_P\sinh^2\big(\frac{\langle\phi\rangle}{\sqrt{6} M_P}\big)\,(\partial_\mu\theta)^2\simeq \frac{1}{2} f^2_a\,(\partial_\mu\theta)^2$, so we can take the canonical axial mode as $a=f_a\,\theta$, which is identified as the axion solving the strong CP problem.

For a general PQ conserving potential, we note that it takes the following form for the canonical radial mode in the Einstein frame,
\bea
V_{\rm PQ}(\phi) =V'_0+3^m\beta_mM^4_P  \bigg[\tanh\Big(\frac{\phi}{\sqrt{6}M_P}\Big)\bigg]^{2m} - 3m^2_\Phi M^2_P  \tanh^2\Big(\frac{\phi}{\sqrt{6}M_P}\Big).
\eea
In this case, for $\phi\ll \sqrt{6} M_P$, we can obtain the VEV of the PQ field in the true vacuum from $2m \beta_m \phi^{2m-1}/(2^m M^{2m-4}_P)=m^2_\Phi\phi$, that is, $\langle\rho\rangle\sim\langle\phi\rangle\sim f_a=(m^2_\Phi M^{2m-4}_P)^{1/(2m-2)}$. For instance, for $m=3$, we get the VEV of the PQ field as $f_a=\sqrt{m_\Phi M_P}$.

\subsection{Axion quality problem}

In the presence of the PQ anomalies, we get the effective gluon couplings violating the strong CP, as follows, 
\bea
{\cal L}_{\rm gluons}=\frac{g^2_s}{32\pi^2} \Big({\bar\theta}+\xi\frac{a}{f_a}\Big) G^a_{\mu\nu}{\tilde G}^{a\mu\nu}
\eea
where $\xi$ is the PQ anomaly coefficient, which is set to $\xi=1$ in KSVZ models.
Then, after the QCD phase transition, there appears an extra contribution to the axion potential, in the following form,
\bea
\Delta V_E =-\Lambda^4_{\rm QCD} \cos\Big({\bar\theta}+\xi\frac{a}{f_a}\Big). \label{QCDpot}
\eea
 After the radial mode settles down to the minimum of the potential, i.e. $\langle\rho\rangle\simeq f_a$, from eqs.~(\ref{Einpot}) and (\ref{QCDpot}), the effective potential for the axion after the QCD phase transition is given by
\bea
V_{\rm eff}(a) =V_0-\Lambda^4_{\rm QCD} \cos\Big({\bar\theta}+\xi\frac{a}{f_a}\Big)+M^4_P\bigg(\frac{f_a}{\sqrt{2}M_P}\bigg)^n \,\sum_{k=0}^{[n/2]} |c_k| \cos\Big((n-2k)\frac{a}{f_a}+A_k\Big). \label{axionpot}
\eea

In order to solve the strong CP problem by the axion, the axion potential needs to relax the effective $\theta$ term dynamically \cite{axionquality} to satisfy the EDM bound,
\bea
|\theta_{\rm eff}|=\bigg|{\bar\theta}+\xi\frac{\langle a\rangle}{f_a}\bigg|<10^{-10}. \label{EDMbound}
\eea
Then, from the minimization of the effective potential for the axion in eq.~(\ref{axionpot}), namely, $\frac{dV_{\rm eff}}{da}=0$, we obtain
\bea
a_{\rm phys}\equiv a+\frac{f_a}{\xi} {\bar\theta} \simeq \frac{f^{n-1}_a}{2^{n/2}M^{n-4}_P m^2_a}\, \sum_{k=0}^{[n/2]} |c_k| (n-2k)\sin\Big( A_k-\frac{n-2k}{\xi}\,{\bar\theta}\Big) \label{axionVEV}
\eea
where $m^2_a=\frac{\xi^2}{f^2_a}\,\Lambda^4_{\rm QCD}$ is the squared mass for the axion due to QCD only, and we assumed $(n-2k) \frac{a_{\rm phys}}{f_a}\ll 1$ and $(n-2k)^2|c_k|f^{n-2}_a/(2^{n/2}M^{n-4}_P)\cos\big( A_k-\frac{n-2k}{\xi}\,{\bar\theta}\big)\lesssim m^2_a$ for all $k$. 

As a result, from eq.~(\ref{axionVEV}) with eq.~(\ref{EDMbound}), we can solve the strong CP problem if
\bea
\frac{\xi f^{n-2}_a}{2^{n/2}M^{n-4}_P m^2_a}\, \sum_{k=0}^{[n/2]} |c_k| (n-2k)\sin\Big( A_k-\frac{n-2k}{\xi}\,{\bar\theta}\Big) < 10^{-10}.
\eea
Unless there is a cancellation between various contributions at the same order, each term in the PQ violating potential is constrained by
\bea
\bigg(\frac{f_a}{M_P}\bigg)^{n}\lesssim \frac{2^{n/2}\xi}{(n-2k)|c_k|} \bigg(\frac{\Lambda_{\rm QCD}}{M_P}\bigg)^4.
\eea 
Then, for a given axion decay constant $f_a$, we can bound the order of the PQ violating polynomial. For instance, for $f_a=10^{12}\,{\rm GeV}$, $\xi=1$ and $|c_k|={\cal O}(1)$, we need $n\gtrsim 12$ for the axion quality.  In general, the lower bound on the order of the PQ violating polynomial scales by $n\propto 1/\ln(M_P/f_a)$. So, for $f_a=10^6\,{\rm GeV}$, we need $n\gtrsim 6$. 

However, considering the bound from the CMB normalization in our model, as will be seen in the later section,  the coefficient of the PQ violating term is bounded by $3^{n/2} |c_k|\lesssim 10^{-10}$ for all $k$. Thus, it is sufficient to take $n\gtrsim 10(5)$ for  $f_a=10^{12}(10^6)\,{\rm GeV}$, $\xi=1$ and $|c_k|\lesssim 10^{-12}$.

\subsection{Axion-photon coupling}

In the presence of charged fermions with nonzero PQ charges, we also obtain the effective axion-photon coupling below the PQ symmetry breaking scale, as follows,
\bea
{\cal L}_{\rm photon} =\frac{1}{4} g_{a\gamma\gamma} \, F_{\mu\nu} {\tilde F}^{\mu\nu}
\eea
with
\bea
g_{a\gamma\gamma} =\frac{\alpha}{2\pi f_a/\xi} \bigg(\frac{E}{N}-1.92\bigg).
\eea
In the minimal scenario where the QCD anomalies for the PQ symmetry are originated from an extra heavy quark with zero electric charge, we get $\xi=1$ and $E/N=0$.

\section{Inflationary dynamics}

 In this section, we discuss the inflationary predictions and bounds in the pole inflation for the PQ field. 
 We divide our discussion into two cases, depending on whether the PQ violating terms are relevant for inflation or not: inflation with PQ conservation and inflation with PQ violation. We note that there were similar proposals for the inflation scenarios based on the PQ field, but with the non-minimal coupling taken to be different from the conformal coupling \cite{PQ0,PQ1,PQ2}.

\subsection{Inflaton equations}

First, from eq.~(\ref{Elag}), we obtain the equation of motion for the radial mode in the following,
\bea
{\ddot\phi}+3H {\dot \phi} -\sqrt{6}M_P\,\sinh\Big(\frac{\phi}{\sqrt{6}M_P}\Big) \cosh\Big(\frac{\phi}{\sqrt{6}M_P}\Big) \, {\dot\theta}^2=-\frac{\partial V_E}{\partial \phi}. \label{eom-phi}
\eea
Thus, for a slow-roll inflation with ${\ddot\phi}\ll H {\dot\phi}$ and ${\dot\theta}\ll H$, we simply get 
\bea
{\dot\phi}\simeq -\frac{1}{3H} \frac{\partial V_E}{\partial \phi}=-\sqrt{2\epsilon_\phi}\, M_P H \label{slow-phi}
\eea
where $\epsilon_\phi$ is the slow-roll parameter for the radial mode, given by $\epsilon_\phi=\frac{M^2_P}{2 V^2_E} \big(\frac{\partial V_E}{\partial\phi}\big)^2$. Here, the Hubble parameter is determined by
\bea
H^2=\frac{1}{3M^2_P}\, \bigg(\frac{1}{2}(\partial_\mu \phi)^2+ 3M^2_P\sinh^2\Big(\frac{\phi}{\sqrt{6} M_P}\Big)\,(\partial_\mu\theta)^2+ V_E\bigg).
\eea
which is approximate to $H^2\simeq \frac{V_E}{3M^2_P}$ during inflation. 

Moreover, from eq.~(\ref{Elag}), we also obtain the equation of motion for the axial mode during inflation as
\bea
6M^2_P\sinh^2\Big(\frac{\phi}{\sqrt{6}M_P}\Big)\bigg[{\ddot\theta}+ 3H {\dot \theta} + \frac{2}{\sqrt{6} M_P} \coth\Big(\frac{\phi}{\sqrt{6}M_P}\Big)\, {\dot\phi}\,{\dot\theta}\bigg] = -\frac{\partial V_E}{\partial\theta}.  \label{eom-theta}
\eea
Then, taking ${\ddot\theta}\ll H {\dot\theta}$ and ${\dot\phi}\ll H$ for a slow-roll inflation, we obtain the approximate equation for the axion kinetic misalignment by
\bea
{\dot\theta}\simeq  -\frac{1}{3H} \, \frac{\frac{\partial V_E}{\partial\theta}}{6M^2_P\sinh^2\big(\frac{\phi}{\sqrt{6}M_P}\big)}=-\frac{\sqrt{2\epsilon_\theta} \,H}{6\sinh^2\big(\frac{\phi}{\sqrt{6}M_P}\big)} \label{slow-theta}
\eea
where $\epsilon_\theta$ is the slow-roll parameter for the angular mode in the second equality, namely, $\epsilon_\theta=\frac{1}{2V^2_E} \big(\frac{\partial V_E}{\partial\theta}\big)^2$. Therefore, as compared to the velocity of the radial mode in eq.~(\ref{slow-phi}), the velocity of the angular mode has an extra suppression factor due to the non-canonical kinetic term during inflation where $\phi\gg \sqrt{6} M_P$. We can set the initial kinetic misalignment of the axion by using eq.~(\ref{slow-theta}) at the end of inflation.

For a slow-roll inflation, we need to take $\rho\sim \sqrt{6} M_P\gg f_a$ and ignore the bare cosmological constant in eq.~(\ref{Einpot}). Then,  the potential for inflation is approximated to 
\bea
V_E(\phi,\theta)\simeq  9 \lambda_\Phi M^4_P\tanh^4\Big(\frac{\phi}{\sqrt{6}M_P}\Big) + V_n(\theta) \tanh^n\Big(\frac{\phi}{\sqrt{6}M_P}\Big),\label{inflatonpot}
\eea
with
\bea
V_n(\theta)\equiv 3^{n/2}M^4_P \sum_{k=0}^{[n/2]} |c_k| \cos\Big((n-2k)\theta+A_k\Big).
\eea

\subsection{Inflation with PQ conservation}

We take the inflaton potential in eq.~(\ref{inflatonpot}) to be dominated by the PQ conserving term in the following,
\bea
V_E(\phi)\simeq V_I \bigg[  \tanh\Big(\frac{\phi}{\sqrt{6}M_P}\Big)\bigg]^{2m}, \label{inflation1}
\eea
with $V_I\equiv 3^m \beta_m M^4_P$.
Then, we first obtain the slow-roll parameters \cite{Clery:2023ptm},
\bea
\epsilon &=&\frac{M^2_P}{2}  \bigg(\frac{1}{V_E}\frac{\partial V_E}{\partial\phi}\bigg)^2 \nonumber \\
&=& \frac{4}{3}m^2  \bigg[  \sinh\Big(\frac{2\phi}{\sqrt{6}M_P}\Big)\bigg]^{-2} ,  \label{ep} \\
\eta &=& \frac{M^2_P}{V_E} \frac{\partial^2V_E}{\partial\phi^2} \nonumber \\
&=&-\frac{4}{3}m \bigg[  \cosh\Big(\frac{2\phi}{\sqrt{6}M_P}\Big)-4 \bigg]  \bigg[\sinh\Big(\frac{2\phi}{\sqrt{6}M_P}\Big)\bigg]^{-2}. \label{eta}
\eea
The number of efoldings is 
\bea
N&=&\frac{1}{M_P} \int^{\phi_*}_{\phi_e} \frac{ {\rm sgn} (V'_E)d\phi}{\sqrt{2\epsilon}} \nonumber \\
&=&\frac{3}{4m}\, \bigg[ \cosh\Big(\frac{2\phi_*}{\sqrt{6}M_P}\Big)- \cosh\Big(\frac{2\phi_e}{\sqrt{6}M_P}\Big)  \bigg] \label{efold}
\eea
where $\phi_*, \phi_e$ are the values of the radial mode at horizon exit and the end of inflation, respectively. Here, we note that $\epsilon=1$ determines $\phi_e$.
As a result, using  eqs.~(\ref{ep}), (\ref{eta}) and (\ref{efold}) and $N\simeq \frac{3}{8}\,  \cosh\Big(\frac{2\phi_*}{\sqrt{6}M_P}\Big)$ for $\phi_*\gg \sqrt{6} M_P$ during inflation, we obtain the slow-roll parameters at horizon exit in terms of the number of efoldings as
\bea
\epsilon_* &\simeq & \frac{3}{4\big(N^2-\frac{9}{16m^2}\big)},  \\
\eta_* &\simeq& \frac{3-2N}{2\big(N^2-\frac{9}{16m^2}\big)}.
\eea
Thus, we get the spectral index in terms of the number of efoldings, as follows, 
\bea
n_s&=&1-6\epsilon_*+2\eta_* \nonumber \\
&=& 1-\frac{4N+3}{2\big(N^2-\frac{9}{16m^2}\big)}. \label{sindex}
\eea
Moreover, the tensor-to-scalar ratio at horizon exit is
\bea
r=16\epsilon_* =  \frac{12}{N^2-\frac{9}{16m^2}}. \label{ratio}
\eea

As a result, from eq.~(\ref{sindex}), we obtain the spectral index as $n_s=0.966$ for $N=60$ being insensitive to $m$, which agrees with the observed spectral index from Planck, $n_s=0.967\pm 0.0037 $ \cite{planck}.
Moreover, we also predict  the tensor-to-scalar ratio as $r=0.0033$ for $N=60$, which is again compatible with  the bound from the combined Planck and Keck data \cite{keck}, $r<0.036$.

The CMB normalization, $A_s=\frac{1}{24\pi^2} \frac{V_I}{\epsilon_* M^4_P}=2.1\times 10^{-9}$,  bounds the quartic coupling for the PQ field by
\bea
3^m \beta_m= (3.1\times 10^{-8}) \,r=1.0\times 10^{-10} \label{CMB}
\eea
where we took $r=0.0033$, in the second equality. For instance, for $m=2$, we have $\beta_m=\lambda_\Phi$, so we need to impose $\lambda_\Phi=1.1\times 10^{-11}$ during inflation. 
Moreover, the inflation with PQ conservation sets the bound on the PQ violating terms, as follows,
\bea
V_n(\theta_i)/M^4_P= 3^{n/2}\sum_{k=0}^{[n/2]} |c_k| \cos\Big((n-2k)\theta_i+A_k\Big) <1.0\times 10^{-10}. \label{PQVbound}
\eea

\subsection{Inflation with PQ violation}

We take the case where the inflaton potential is dominated by the PQ violating terms in the following,
\bea
V_E(\phi,\theta)\simeq V_n(\theta) \cdot W_n(\phi) . \label{inflation2}
\eea 
with
\bea
W_n(\phi)\equiv  \bigg[\tanh\Big(\frac{\phi}{\sqrt{6}M_P}\Big)\bigg]^n.
\eea
In this case, as far as the inflaton potential is positive definite during inflation, both the radial and angular modes of the PQ field could serve as the inflaton, corresponding to a multi-field inflation scenario \cite{Gong:2011cd}. Generically, the angular mode should not be located at the minimum of the potential during inflation. The reason is the following. For instance, taking $A_k$ in eq.~(\ref{inflatonpot}) to be the same for all $k$, $V_n(\theta)$ becomes always negative at the minimum of the potential, so it would be not appropriate for inflation with PQ violation. 

Even if the PQ violating terms can be dominant during inflation, they become sub-dominant as the PQ field evolves after inflation, such that the PQ conserving term takes over in determining the dynamics of the radial mode of the PQ field. Then, from the initial condition that  the axial mode is deviated from the minimum during inflation, the axion can settle down safely toward the minimum of the potential after inflation. 

For inflation with PQ violation, the slow-roll parameters are determined from the derivatives of the PQ violating terms in the potential, so $\epsilon_\phi\simeq \frac{M^2_P}{2W^2_n} \big(\frac{\partial W_n}{\partial\phi}\big)^2$ and $\epsilon_\theta\simeq \frac{1}{2V^2_n} \big(\frac{\partial V_n}{\partial\theta}\big)^2$.  In this case, the slow-roll parameter for the radial mode is still small for $\phi\gg \sqrt{6} M_P$, but the slow-roll parameter for the angular mode is of order one. Nonetheless, the velocity of the angular mode is sufficiently small for $\phi\gg \sqrt{6} M_P$ due to its non-canonical kinetic term as can be seen in eq.~(\ref{slow-theta}). For generality, we can introduce the effective slow-roll parameter by taking into account the non-canonical kinetic term \cite{Gong:2011cd}, as follows,
\bea
\epsilon=g^{\phi\phi}\epsilon_\phi + g^{\theta\theta} \epsilon_\theta \label{genslow}
\eea
with $g^{\phi\phi}=1$ and $g^{\theta\theta}=1/\big[6M^2_P\sinh^2\big(\frac{\phi}{\sqrt{6}M_P}\big)\big]$.

When the inflation energy comes dominantly from the PQ violating terms and the axion field value is set to $\theta=\theta_i$ during inflation, we obtain the constraint from the CMB normalization as
\bea
V_n(\theta_i)= 3^{n/2}\sum_{k=0}^{[n/2]} |c_k| \cos\Big((n-2k)\theta_i+A_k\Big) =(3.1\times 10^{-8}) \,r
\eea
where $r=16\epsilon_*$ with $\epsilon_*$ being evaluated from eq.~(\ref{genslow}) at the horizon exit.

\subsection{Post-inflationary evolution of the fields}

After inflation, the radial mode takes a sub-Planckian field value with $\phi\ll \sqrt{6} M_P$ so we can approximate the equations of motion in eqs.~(\ref{eom-phi}) and (\ref{eom-theta}), as follows,
\bea
{\ddot\phi}+3H {\dot \phi} -\phi \, {\dot\theta}^2&\simeq& -\frac{\partial V_E}{\partial \phi},  \label{approx-phi} \\
\phi^2 ({\ddot\theta}+ 3H {\dot \theta}) + 2\phi {\dot\phi}\,{\dot\theta}&\simeq&  -\frac{\partial V_E}{\partial\theta} \label{approx-theta}
\eea
where the Einstein-frame potential in eq.~(\ref{Einpot}) becomes 
\bea
V_E\simeq V_0+\frac{1}{4} \lambda_\Phi (\phi^2-f^2_a)^2 + \frac{1}{2^{n/2}} \bigg(\frac{\phi}{M_P}\bigg)^n M^4_P  \sum_{k=0}^{[n/2]} |c_k| \cos\Big((n-2k)\theta+A_k\Big).
\eea

\begin{figure}[!t]
\begin{center}
 \includegraphics[width=0.45\textwidth,clip]{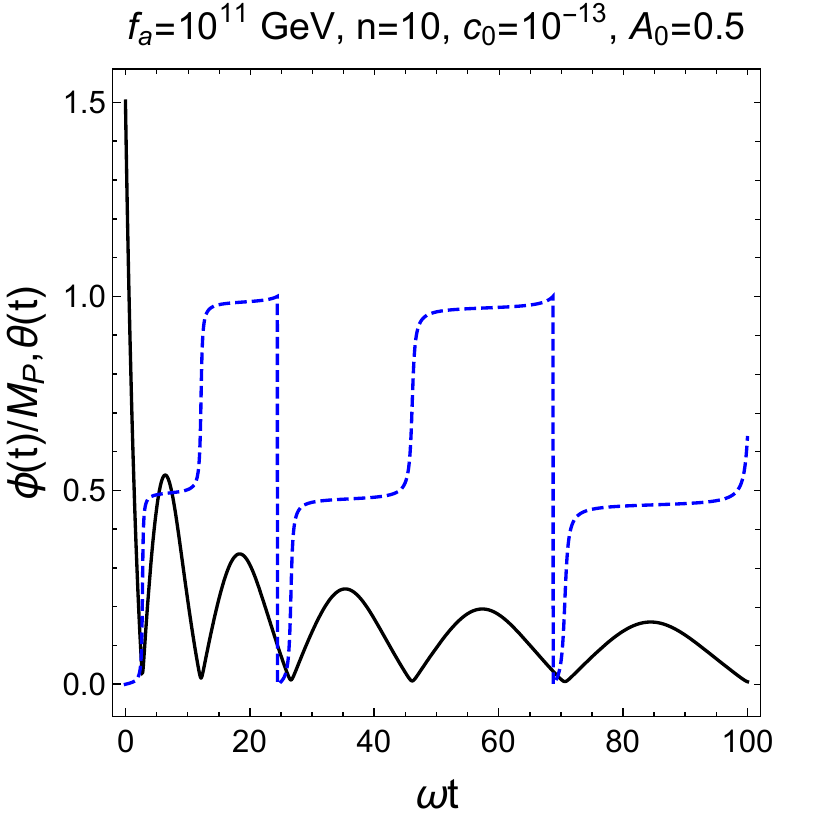}\,\,  \includegraphics[width=0.45\textwidth,clip]{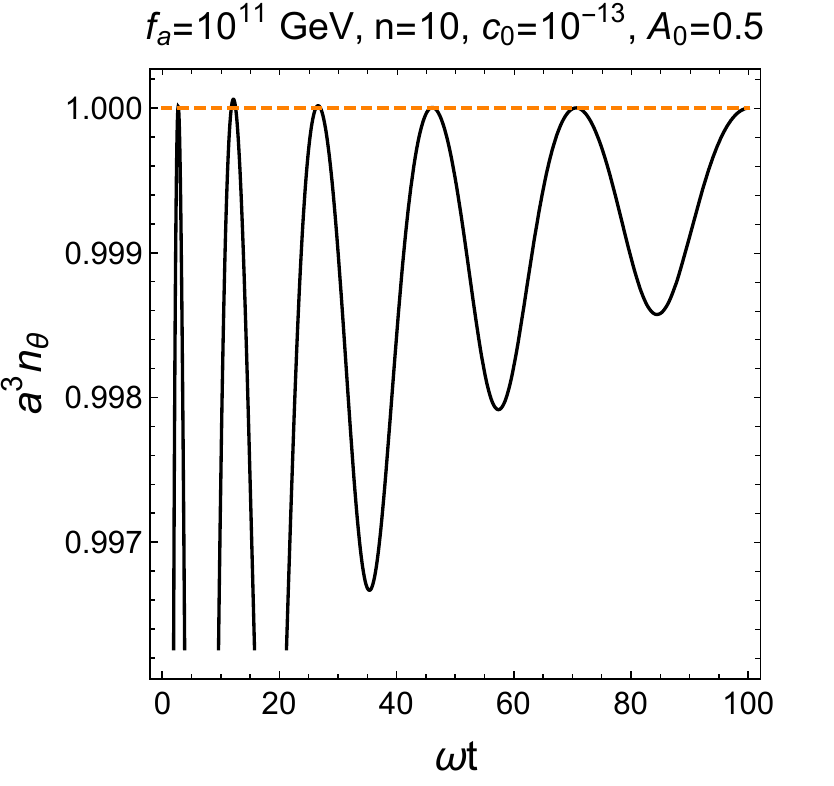}  
 \end{center}
\caption{(Left) Post-inflationary evolution of the PQ inflaton and the angular mode as a function of $\omega t$ in black solid and blue dashed lines,  respectively, with $\omega$ being the oscillation frequency of the inflaton condensate. (Right) Post-inflationary evolution of the total PQ Noether charge, $n_\theta=\phi^2 {\dot\theta}$, as a function of $\omega t$. We considered the quartic term for the PQ-invariant potential and took $f_a=10^{11}\,{\rm GeV}$ and the parameters in the PQ violating potential to $n=10$, $c_0=10^{-13}$, $A_0=0.5$, and $c_k=0$ for $k\neq 0$.}
\label{fig:oscillation}
\end{figure}

We consider the case where the PQ conserving term is dominant for inflation. Then, as the amplitude of the radial mode decreases until reaching the VEV, the PQ violating terms become less important. When the PQ violating terms are sufficiently small during the evolution of the fields, we have the total Noether charge for the PQ symmetry \cite{Co:2019jts} approximately conserved after inflation. Namely, for $n_\theta=\phi^2 {\dot\theta}$, we get $\frac{d}{dt}(a^3 n_\theta)=0$ from  eq.~(\ref{approx-theta}). Thus, we can approximate the equation of motion for the radial mode in eq.~(\ref{approx-phi}) for $f_a\lesssim \phi\ll M_P$ as
\bea
{\ddot\phi}+3H {\dot \phi} \simeq \frac{C^2}{a^6 \phi^3} -\lambda_\Phi \phi^3
\eea
where $C=a_e^3 \phi^2_e{\dot\theta}_e$ is the integration of constant given at the end of inflation. Therefore, the centrifugal force term due to the angular motion decreases rapidly as the universe is expanding, so the radial mode undergoes a coherent motion around the vacuum  dominantly by the quartic term during reheating.

In Fig.~\ref{fig:oscillation}, considering the PQ invariant quartic potential, we depict the time evolution of the PQ inflaton and the angular mode on left and the total PQ Noether charge on right as a function of $\omega t$.  We took $f_a=10^{11}\,{\rm GeV}$, and the parameters in the PQ violating potential to $n=10$, $c_0=10^{-13}$, $A_0=0.5$, and $c_k=0$ for $k\neq 0$. We show that the PQ inflaton undergoes a damped oscillation dominated by the PQ invariant potential from the left figure, and the total PQ Noether charge oscillates but it approaches to the constant value, which is set by the initial condition during inflation.   

We also remark on the case where PQ violating terms are dominant for inflation. In this case, the PQ violating terms are still important in the post-inflationary dynamics, so the total Noether charge for the PQ symmetry evolves nontrivially during reheating, so we need to solve two coupled equations of motion in eqs.~(\ref{approx-phi}) and (\ref{approx-theta}), with $V_E\simeq V_n(\theta) \cdot W_n(\phi)$, for the post-inflationary evolution of the PQ field.

\section{Reheating}

We consider the inflaton condensate and the general equation of state during reheating.
Then, we first discuss the perturbative reheating from the decay or scattering of the PQ inflaton into the extra fermions and the Higgs and determine the minimal reheating temperature.

\subsection{Inflaton condensate}

Suppose that the reheating dynamics is dominated by the radial mode of the PQ field.
Then, for $|\Phi|\ll \sqrt{3}M_P$, we can approximate the kinetic term for the radial mode of the PQ field in Einstein frame in a canonical form, that is, $\phi\simeq \rho$, so the Einstein-frame potential for the pole inflation in eqs.~(\ref{inflation1}) takes
\bea
V_E(\phi)\simeq \alpha_m \phi^{2m}, \label{rehpot}
\eea
with $\alpha_m=\beta_m M^{4-2m}_P/2^m$.  For $m=2$, we get $\beta_m=\lambda_\Phi$. 
Thus, as the inflaton potential becomes anharmonic for $m>1$ during reheating,  we get a general equation of state during reheating. 
After the period of exponential expansion, the inflaton $\phi$ begins to oscillate about the minimum of the potential. 
For the inflaton potential given in eq.~(\ref{inflation1}), the inflation field value at the end of inflation is determined by $\ddot a=0$ \cite{Clery:2023ptm} to be
\begin{align}
   \phi_{\rm end} &\simeq 
   \sqrt{\frac{3}{8}}M_P\ln\left[\frac{1}{2} + \frac{2m}{3}\left(2m+ \sqrt{4m^2+3}\right)\right]. 
   \label{phiend}
\end{align}
The condition $\ddot a=0$ is equivalent to $\dot\phi_{\rm end}^2 = V_E(\phi_{\rm end})$, so
 the inflaton energy density at $\phi_{\rm end}$ is  $\rho_{\rm end}=\frac{3}{2}V_E(\phi_{\rm end})$.

The averaged energy density and pressure for the inflaton becomes
\bea
\rho_\phi &=&\Big\langle \frac{1}{2}{\dot\phi}^2\Big\rangle+\langle V_E(\phi)\rangle=(m+1) \langle V_E(\phi)\rangle, \label{infdens} \\
p_\phi &=& \Big\langle \frac{1}{2}{\dot\phi}^2\Big\rangle-\langle V_E(\phi)\rangle=(m-1) \langle V_E(\phi)\rangle.
\eea
Then, the averaged equation of state for the inflaton during reheating is given by
\bea
\langle w_\phi\rangle=\frac{p_\phi}{\rho_\phi} =\frac{m-1}{m+1}.  \label{eos1}
\eea
For the PQ conserving potential with $m=2$, we get $\langle w_\phi\rangle=\frac{1}{3}$ for $m=2$, so it is the same as the one for radiation. 

We take the inflaton to be $\phi=\phi_0(t) {\cal P}(t)$, where $\phi_0(t)$ is the amplitude of the inflaton oscillation and it is constant over one oscillation, and  ${\cal P}(t)$ is the periodic function.  From eq.~(\ref{infdens}) with the energy conservation, we obtain $\rho_\phi=(m+1) \langle V_E(\phi)\rangle=V_E(\phi_0)$. Then, we get $\langle {\cal P}^{2m}\rangle=\frac{1}{m+1}$. 

From the energy density for the inflaton,
\bea
 \frac{1}{2}{\dot\phi}^2+V_E(\phi)=V_E(\phi_0),
\eea
we obtain the equation for the periodic function ${\cal P}$ as follows,
\bea
{\dot{\cal P}}^2 =\frac{2\rho_\phi}{\phi^2_0} \Big(1-{\cal P}^{2m}\Big)= \frac{m^2_\phi}{m(2m-1)}\,  \Big(1-{\cal P}^{2m}\Big) \label{Peq}
\eea
where we used the effective inflaton mass in the second equality,
\bea
m^2_\phi =V^{\prime\prime}_E(\phi_0) =2\alpha_m  m (2m-1) \phi^{2m-2}_0. \label{inflatonmass}
\eea
Thus, from the integral of  eq.~(\ref{Peq}), we get the angular frequency of the inflaton oscillation \cite{Garcia:2020wiy,Clery:2023ptm} as
\bea
\omega = m_\phi\sqrt{\frac{\pi m}{2m-1}}\, \frac{\Gamma\big(\frac{1}{2}+\frac{1}{2m}\big)}{\Gamma\big(\frac{1}{2m}\big)}. \label{freq}
\eea
As a result, we can make a Fourier expansion of the periodic function  $\cal P$ by
\bea
{\cal P}(t) =\sum_{n=-\infty}^\infty {\cal P}_n \,e^{-in\omega t}. 
\eea

\subsection{Boltzmann equations for reheating}

Including the effects of the Hubble friction and the inflaton decay/scattering, we find the equation of motion for the inflaton, as follows,
\bea
{\ddot\phi}+ (3H+\Gamma_\phi) {\dot\phi} +V'_E=0
\eea
where $\Gamma_\phi$  is the inflaton decay or scattering rate, given by
\bea
\Gamma_\phi=\sum_f\Gamma_{\phi\to f{\bar f}} + \Gamma_{\phi\phi\to HH}+\Gamma_{\phi\phi\to aa}.
\eea
The above equation can be approximated to the Boltzmann equation for the averaged energy density,
\bea
{\dot\rho}_\phi + 3(1+w_\phi)H \rho_\phi\simeq -\Gamma_\phi (1+w_\phi) \rho_\phi.
\label{Boltzmann_phi}
\eea
Moreover, the Boltzmann equation governing the radiation energy density $\rho_R$ is given by
\bea
{\dot\rho}_R + 4 H\rho_R =\Gamma_\phi (1+w_\phi) \rho_\phi. \label{Boltzmann_rad}
\eea

First, we recall that the PQ inflaton has a derivative coupling to the angular mode, ${\cal L}_{\rm int}=\frac{1}{2} \phi^2(\partial_\mu\theta)^2$.
During the oscillation of the radial mode, $\phi$, however, the angular mode has a rapidly changing kinetic term, so it is not clear to see how to identify the axion quanta produced from the scattering of the radial mode in this basis. So, instead, we choose a Cartesian basis for the PQ field by $\Phi=\frac{1}{\sqrt{2}}(\phi+ia)$, for which the kinetic term for the PQ field during reheating takes a canonical form, ${\cal L}_{\rm kin}=\frac{1}{2}(\partial_\mu\phi)^2 +\frac{1}{2}(\partial_\mu a)^2$, and the PQ invariant potential gives rise to the interaction term between two real scalar fields, ${\cal L}_{\rm int}=-\frac{1}{2}\lambda_\Phi \phi^2 a^2$, where the inflaton condensate is $\phi=\phi_0(t) {\cal P}(t)$.  Then, over one oscillation for which  the PQ field oscillates with a large radial distance, we can take the real part to be the radial mode and consider the imaginary part as being along the orthogonal direction to the radial mode\footnote{In the vacuum, we note that  the axion in the leading order expansion of the field in the polar form gives rise to $\Phi=\frac{1}{\sqrt{2}}f_a\,e^{i\theta}\simeq \frac{1}{\sqrt{2}}f_a(1+i\theta)=\frac{1}{\sqrt{2}}(f_a+ia)$, which is identical to the cartesian form of the PQ field.}. Thus, the Cartesian form is  appropriate for describing the axion quanta from the inflaton scattering.
As a result, we get the scattering rate of the inflaton condensate \cite{Garcia:2020wiy,Clery:2023ptm} for $\phi\phi\to aa$, as follows, 
\bea
\Gamma_{\phi\phi\to aa}=\frac{1}{8\pi (1+w_\phi)\rho_\phi} \sum_{n=1}^\infty |M^a_n|^2 (E_n\beta^{a}_n),
\eea
with 
\bea
|M^a_n|^2 =  4\lambda^2_\Phi \phi^4_0 |({\cal P}^2)_n|^2,
\eea
and $\beta^{a}_n= \sqrt{1-\frac{m^2_{a}}{E^2_n}}$.
Here, for $\phi\gg f_a$, $m^2_{a}=\lambda_\Phi \phi^2$  which is the effective mass of the axion during reheating \footnote{As the inflaton condensate settles down close to the minimum of the potential, the quadratic term in the PQ invariant potential becomes relevant. Then, the effective mass of the axion is corrected to $m^2_{a}=\lambda_\Phi \phi^2-m^2_\Phi$, which vanishes in the vacuum with $\langle\phi\rangle=f_a$, up to the PQ violating terms.}. From eq.~(\ref{freq}), we get $\omega^2=0.72\lambda_\Phi \phi^2_0$ for $m=2$, so $2\omega> m_a$, making $\phi\phi\to aa$ open kinematically. We also note that $({\cal P}^2)_n$ are the Fourier coefficients of the expansion, ${\cal P}^2=\sum_{n=-\infty}^\infty ({\cal P}^2)_n e^{-in\omega t}$. For the PQ invariant potential with $m=2$, the first few nonzero coefficients are given by $2({\cal P}^2)_2=0.4972, 2({\cal P}^2)_4=0.04289, 2({\cal P}^2)_6=0.002778$, etc. Thus, the corresponding averaged scattering rate for the inflaton is given by
\bea
\langle \Gamma_{\phi\phi\to aa}\rangle =\frac{\lambda^2_\Phi \phi^2_0 \omega}{2\pi m^2_\phi}\,(m+1)(2m-1) \Sigma^a_m \bigg\langle \bigg(1-\frac{m^2_a}{\omega^2n^2}\bigg)^{1/2}\bigg\rangle \label{scatteringrate1}
\eea
with $\Sigma^a_m=\sum^\infty_{n=1}n |({\cal P}^2)_n|^2$.

Similarly, the Higgs-portal interactions of the PQ inflaton, ${\cal L}_{\rm H-portal}=-\lambda_{H\Phi}|\Phi|^2|H|^2$, give rise to the scattering rate of the inflaton condensate, as follows,
\bea
\Gamma_{\phi\phi\to HH}=\frac{1}{8\pi (1+w_\phi)\rho_\phi} \sum_{n=1}^\infty |M^H_n|^2 (E_n\beta^H_n),
\eea
with 
\bea
|M^H_n|^2 = 4\lambda_{H\Phi}^2 \phi^4_0|({\cal P}^2)_n|^2
\eea
and $\beta^H_n = \sqrt{1-\frac{m^2_H}{E^2_n}}$. Here, the effective masses for the Higgs fields are given by $m^2_H=m^2_{H,0}+\frac{1}{2}\lambda_{H\Phi} \phi^2(t)$ where $m^2_{H,0}$ is the bare Higgs mass parameter.  Thus, the corresponding averaged scattering rate for the inflaton is given by
\bea
\langle \Gamma_{\phi\phi\to HH}\rangle =\frac{\lambda^2_{H\Phi}\phi^2_0 \omega}{\pi m^2_\phi}\,(m+1)(2m-1) \Sigma^H_m \bigg\langle \bigg(1-\frac{m^2_H}{\omega^2n^2}\bigg)^{1/2}\bigg\rangle \label{scatteringrate2}
\eea
with $\Sigma^H_m=\sum^\infty_{n=1}n |({\cal P}^2)_n|^2$. 
We remark that we need to take the Higgs-portal coupling to be small enough in order to keep the running quartic coupling $\lambda_\Phi$ at the order of $10^{-11}$, namely, $|\lambda_{H\Phi}|\lesssim 10^{-5}$.

On the other hand, in order to solve the strong CP problem by the anomalous couplings of the axion to gluons, it is necessary to introduce the Yukawa couplings to the extra fermions such as extra heavy quarks $Q$ in KSVZ axion models. Moreover, the PQ inflaton can be responsible for the generation of masses for the right-handed neutrinos, $N_i(i=1,2,3)$, through the Yukawa couplings. Thus, we consider the Yukawa couplings as ${\cal L}_{\rm int}=-\frac{1}{\sqrt{2}}y_f\phi {\bar f}_L f_R+{\rm h.c.}$, where $f_L=Q_L$ and $f_R=Q_R$ for extra heavy quarks or $f_L=N^c_{i,L}$ and $f_R=N_{i,R}$ for right-handed neutrinos. Then, we obtain the decay rate of the inflaton condensate, as follows,
\bea
\Gamma_{\phi\to f{\bar f}}=\frac{1}{8\pi (1+w_\phi)\rho_\phi} \sum_{n=1}^\infty |M^f_n|^2 (E_n\beta^f_n)
\eea
where $E_n=n \omega$,
\bea
|M^f_n|^2=N_c y^2_f \phi^2_0 |{\cal P}_n|^2 E^2_n (\beta^f_n)^2,
\eea
with $N_c$ being the number of colors for the extra fermion $f$, $\beta^f_n = \sqrt{1-\frac{4m^2_f}{E^2_n}}$, and ${\cal P}_n$ are the Fourier coefficients of the expansion, ${\cal P}=\sum_{n=-\infty}^\infty {\cal P}_n e^{-in\omega t}$. For the PQ-invariant potential with $m=2$, the first few nonzero coefficients are given by $2{\cal P}_1=0.9550, 2{\cal P}_3=0.04305, 2{\cal P}_5=0.001859$, etc.
Then, averaging over oscillations, we get
\bea
\langle\Gamma_{\phi\to f{\bar f}}\rangle&=&\frac{N_c y^2_f \phi^2_0 \omega^3}{8\pi (1+w_\phi)\rho_\phi}\, \sum_{n=1}^\infty  n^3 |{\cal P}_n|^2 \langle\beta^3_n\rangle \nonumber \\
&=& \frac{N_c y^2_f \omega^3}{8\pi m^2_\phi}\,(m+1)(2m-1) \Sigma^f_m \Bigg\langle\bigg(1-\frac{4m^2_f}{\omega^2 n^2}\bigg)^{3/2}\Bigg\rangle \label{decayrate}
\eea
with $\Sigma^f_m =\sum_{n=1}^\infty  n^3 |{\cal P}_n|^2$.
Here, if the extra fermions receive masses only from the Yukawa couplings to the PQ inflaton, their masses are given by $m_f=\frac{1}{\sqrt{2}} y_f\phi(t)=(m_{f,0}/f_a)\phi(t)$ where $f_a$ is the VEV of the PQ field and $m_{f,0}$ is the fermion mass in the true vacuum. If electroweak symmetry is unbroken during reheating, there is no mixing between the PQ and Higgs fields, so there is no decay process of the PQ inflaton into the SM fermions.  We also remark that the Yukawa couplings to the PQ field must be chosen to be small enough in order to make the running effects on the quartic coupling $\lambda_\Phi$ ignorable, namely, $y_f\lesssim 10^{-3}$. 

The same Yukawa couplings of the PQ field to the extra fermions also give rise to the inflaton scattering, $\phi\phi\to f{\bar f}$, with the corresponding scattering rate,
\bea
\Gamma_{\phi\phi\to f{\bar f}}=\frac{1}{8\pi (1+w_\phi)\rho_\phi} \sum_{n=1}^\infty |{\hat M}^f_n|^2 (E_n {\hat \beta}^f_n)
\eea
where 
\bea
 |{\hat M}^f_n|^2=\frac{32N_c y^4_f m^2_f \phi^4_0 ({\hat \beta}^f_n)^2}{E^2_n}
\eea
and ${\hat \beta}^f_n=\sqrt{1-\frac{m^2_f}{E^2_n}}$. Thus, the averaged scattering rate is given by
\bea
\langle \Gamma_{\phi\phi\to f{\bar f}}\rangle &=& \frac{4N_c y^4_f  \phi^4_0 }{\pi (1+w_\phi)\rho_\phi \omega}\, \sum_{n=1}^\infty  n^{-1}|({\cal P}^2)_n|^2\langle m^2_f ({\hat \beta}^f_n)^3 \rangle \nonumber \\
&=&\frac{4N_c y^4_f \phi^2_0 }{\pi m^2_\phi \omega}\, (m+1)(2m-1)\,{\hat\Sigma}^f_m\bigg\langle  m^2_f \bigg(1-\frac{m^2_f}{n^2\omega^2}\bigg)^{3/2}\bigg\rangle,
\eea
with $ {\hat\Sigma}^f_m=\sum_{n=1}^\infty n^{-1}|({\cal P}^2)_n|^2$.

The decay and scattering rates of the inflaton scale with the inflaton energy density by $\Gamma_{\phi \to f{\bar f}}=\gamma_1 \rho^l_\phi$ with $l=\frac{1}{2}-\frac{1}{2m}$ and $\Gamma_{\phi\phi\to HH}=\gamma_2 \rho^n_\phi$ with $n=\frac{3}{2m}-\frac{1}{2}$ \cite{Garcia:2020wiy,Clery:2023ptm}. Thus, taking $m=2$ in our case, we obtain $l=n=\frac{1}{4}$, so both the decay and scattering rates are comparable.
However, for $m>2$, we have $l>n$, so the decay rate becomes suppressed as compared to the scattering rates.

\subsection{Reheating temperature}

For $a_{\rm end} \ll a\ll a_{\rm RH}$ where $a_{\rm RH}$ is the scale factor at the time reheating is complete, we can ignore the inflaton decay/scattering rates and integrate  eq.~(\ref{Boltzmann_phi}) approximately for $m=2$ to obtain
\be
\rho_{\rm \phi}(a)\simeq \rho_{\rm end}\left(\frac{a_{\rm end}}{a}\right)^{4}. \label{inflatondensity}
\ee
This is due to the equation of state with $\langle w_\phi\rangle=\frac{1}{3}$ during reheating, given in eq.~(\ref{eos1}).
When the reheating process is dominated by the perturbative decays and scattering processes of the inflaton, we obtain the reheating  temperature during reheating, as follows \cite{Clery:2023ptm},
\bea
T_{\rm RH}= \bigg(\frac{30}{\pi g_*(T_{\rm RH})}\bigg)^{1/4} \Big( \frac{4}{3} \sqrt{3}M_P \gamma_\phi\Big).
\eea
where we took the sum of the decay and scattering rates of the inflaton from eqs.~(\ref{decayrate}), (\ref{scatteringrate1}) and (\ref{scatteringrate2}) by $\gamma_\phi\equiv (\sum_f \Gamma_{\phi\to f{\bar f}}+ \Gamma_{\phi\phi\to HH}+\Gamma_{\phi\phi\to f{\bar f}})/\rho^\frac{1}{4}_\phi$.

For a general PQ invariant potential, the scaling of the inflaton energy density is generalized to
\bea
\rho_{\rm \phi}(a)\simeq \rho_{\rm end}\left(\frac{a_{\rm end}}{a}\right)^{\frac{6m}{m+1}}, 
\eea
so we obtain the corresponding reheating temperature \cite{Clery:2023ptm} by
\bea
T_{\rm RH}= \left\{\begin{array}{cc} \bigg(\frac{30}{\pi g_*(T_{\rm RH})}\bigg)^{1/4} \bigg[ \frac{2m}{4+m-6mk}\,(\sqrt{3}M^{2(1-2k)}_P\gamma_\phi)\bigg]^{\frac{1}{2(1-2k)}}, \qquad 4+m-6m k>0, \vspace{0.3cm}\\  \bigg(\frac{30}{\pi g_*(T_{\rm RH})}\bigg)^{1/4}\bigg[\frac{2m}{6mk-4-m}\,(\sqrt{3}M^{2(1-2k)}_P\gamma_\phi)\,( \rho_{\rm end})^{\frac{6mk-m-4}{6m}}\bigg]^{\frac{3m}{4(m-2)}}, \quad 4+m-6mk<0. \end{array} \right.
\eea
Here, we parametrized the decay or scattering rates of the inflaton as $\Gamma_\phi=\gamma_\phi \rho^k_\phi/M^{4k-1}_P$.

Taking the case with $m=2$, we find that 
\bea
\gamma_\phi|_{\rm decay} &\simeq &\frac{3\sqrt{\pi}N_c}{2}\, y^2_f \lambda^{1/4}_\Phi\,\bigg(\frac{\Gamma\big(\frac{3}{4}\big)}{\Gamma\big(\frac{1}{4}\big)}\bigg)^3\,(0.5\Sigma^f_2 {\cal R}^{-1/2}_f), \\
\gamma_\phi|_{\rm scattering} &\simeq & \frac{6}{\sqrt{\pi}}\bigg(\frac{\Gamma\big(\frac{3}{4}\big)}{\Gamma\big(\frac{1}{4}\big)}\bigg) {\rm max}\bigg[\frac{\lambda^2_{H\Phi}}{\lambda^{3/4}_\Phi}\, \,\Sigma^H_2,  \frac{4N_cy^4_f}{\lambda^{3/4}_\Phi}\, \frac{m^2_f}{\omega^2}\,({\hat\Sigma}^f_2 {\hat {\cal R}}_f^{-1/2}) \bigg],
\eea
for  the inflaton decay and the inflaton scattering, respectively. Here, $\Sigma^f_2=0.2406$, $\Sigma^H_2=0.1255$, ${\hat\Sigma}^f_2=0.2282$, and we approximated the averaged phase space factor for $2m_f\gg w$ by ${\cal R}_f\equiv 4m^2_f/w^2$ and ${\hat {\cal R}}_f\equiv m^2_f/w^2$ \cite{Garcia:2020wiy}.
Thus, we can determine the reheating temperature by the inflaton decay into a pair of extra heavy quarks or the inflaton scattering into a pair of the SM Higgs bosons, respectively, as follows,
\bea
T_{\rm RH}|_{\rm decay} &\simeq & 2.9\times 10^4\,{\rm GeV} \bigg(\frac{100}{g_*(T_{\rm reh})}\bigg)^{1/4}\bigg(\frac{y_f}{10^{-4}}\bigg)\bigg(\frac{\lambda_\Phi}{10^{-11}}\bigg)^{1/4}, \label{decay}  \\
T_{\rm RH}|_{\rm scattering} &\simeq& 6.0\times 10^{11}\,{\rm GeV}  \bigg(\frac{100}{g_*(T_{\rm reh})}\bigg)^{1/4} \bigg(\frac{{\rm max}[\lambda_{H\Phi},\sqrt{4N_c}y^2_f m_f/\omega]}{10^{-7}}\bigg)^2 \bigg(\frac{10^{-11}}{\lambda_\Phi}\bigg)^{3/4}. \label{scattering}
\eea
Therefore, we find that the inflaton scattering with the Higgs-portal coupling is more efficient for reheating.

\subsection{Dark radiation from axions}

Taking the ratio of the scattering rates in eqs.~(\ref{scatteringrate1}) and (\ref{scatteringrate2}), we obtain
\bea
\frac{\Gamma_{\phi\phi\to aa}}{\Gamma_{\phi\phi\to HH}}\simeq \frac{\lambda^2_\Phi}{2\lambda^2_{H\Phi}}.
\eea 
As a result, for $\lambda_{H\Phi}\gtrsim \frac{1}{\sqrt{2}}\lambda_\Phi$, the inflaton scattering is dominated by $\phi\phi\to HH$. However, the produced axions can contribute to the effective number of neutrinos, $\Delta N_{\rm eff}$. 

When the produced axions remain out of equilibrium after reheating, the correction to the effective number of neutrino species is given as follows \cite{thermalaxions2},
\bea
\Delta N_{\rm eff}= 0.02678\, \bigg(\frac{Y_a}{Y^{\rm eq}_a}\bigg) \bigg(\frac{106.75}{g_{*s}(T_{\rm reh})}\bigg)^{4/3}
\eea
where $Y_a$ is the axion abundance produced from the inflaton scattering, $Y^{\rm eq}_a$ is the axion abundance at equilibrium given by $Y^{\rm eq}_a=\frac{45 \zeta(3)}{2\pi^4 g_{*s}(T_{\rm reh})}$, and we took $N^{\rm SM}_{\rm eff}=3.0440$ in the SM \cite{Neff}. In this case,  for $Y_a\gtrsim 10 Y^{\rm eq}_a$, the excess in the effective number of neutrinos would be in a tension with the current bounds from the Planck satellite, $N_{\rm eff}=2.99\pm 0.17$ \cite{planck2}.

As discussed in the previous subsection, the $\phi\phi\to HH$ scattering is dominant for $\lambda_{H\Phi}\gtrsim \sqrt{4N_c}y^2_f m_f/\omega$. In this case, we can compute the number density of the axions produced from $\phi\phi\to aa$, as follows,
\bea
n_a\simeq {\rm BR}(\phi\phi\to aa)\,\frac{\rho_\phi}{\omega}
\eea
where ${\rm BR}(\phi\phi\to aa)=\Gamma_{\phi\phi\to aa}/(\Gamma_{\phi\phi\to aa}+\Gamma_{\phi\phi\to HH})\simeq \lambda^2_\Phi/(2\lambda^2_{H\Phi})$ for $\lambda_{H\Phi}\gtrsim \frac{1}{\sqrt{2}}\lambda_\Phi$, and we approximated the inflaton to the first Fourier mode with $E=\omega$, ignoring the suppressed higher Fourier modes. 
Thus, from $\rho_\phi=\rho_R$ at reheating completion, we obtain the axion abundance, $Y_a= \frac{n_a}{s} $, at reheating as
\bea
Y_a=  {\rm BR}(\phi\phi\to aa)\,\frac{\rho_\phi}{\omega s} = {\rm BR}(\phi\phi\to aa)\,\frac{\rho_R}{\omega s} ={\rm BR}(\phi\phi\to aa)\, \frac{T_{\rm reh}}{4\omega}.
\eea
Then, using $\rho_R/s=3T_{\rm reh}/4$ and $\omega\simeq  0.85 \lambda^{1/2}_\Phi \phi_0$  for $m=2$, we get
\bea
\frac{Y_a}{Y^{\rm eq}_a} =\frac{3 \lambda^{3/2}_\Phi g_{*s}(T_{\rm reh})T_{\rm reh}}{2.2\lambda^2_{H\Phi}  \phi_0}. \label{axionprod}
\eea
Therefore, from $\phi_0\simeq 1.5M_P$ and $\lambda_\Phi=1.1\times 10^{-11}$, we obtain $Y_a\gtrsim Y^{\rm eq}_a$, provided that 
\bea
T_{\rm reh}\gtrsim 6.8\times 10^{10}\,{\rm GeV} \,\bigg(\frac{\lambda_{H\Phi}}{10^{-11}}\bigg)^2.
\eea

However,  the axions would become thermalized with the SM plasma at a sufficiently high reheating temperature \cite{thermalaxions1,thermalaxions2}, 
\bea
T_{\rm reh}\gtrsim 1.7\times 10^9\,{\rm GeV}\bigg(\frac{f_a}{10^{11}\,{\rm GeV}}\bigg)^{2.246}\equiv T_{a, {\rm eq}}. \label{axionthermal}
\eea
Then, for $T_{\rm reh}>T_{a, {\rm eq}}$, we can compute the contribution of the axions to the effective number of neutrino species just from the abundance in thermal equilibrium $Y^{\rm eq}_a$, as follows,
\bea
\Delta N_{\rm eff}=\frac{4}{7}\bigg(\frac{T_{a,0}}{T_{\nu,0}}\bigg)^4=\frac{4}{7}\bigg(\frac{11}{4}\bigg)^{4/3} \bigg(\frac{g_{*s}(T_0)}{g_{*s}(T_{a, {\rm eq}})}\bigg)^{4/3} \label{Neff}
\eea
where $T_{\nu,0}, T_{a,0}$ are the neutrino and axion temperatures, respectively, at present, and $g_{*s}(T_0)=3.91$. Thus, we get $\Delta N_{\rm eff}=0.02678$ for  $g_{*s}(T_{a, {\rm eq}})=106.75$; $\Delta N_{\rm eff}=0.02229$ for $g_{*s}(T_{a, {\rm eq}})=122.5$ (adding one charge-neutral heavy quark and three right-handed neutrinos to the SM); $\Delta N_{\rm eff}=0.02363$ for $g_{*s}(T_{a, {\rm eq}})=117.25$ (adding one charge-neutral heavy quark to the SM). 
Therefore, the excess in the effective number of neutrinos can be tested in the future CMB experiments such as CMB-S4 \cite{CMBS4}.

Finally, for $T_{\rm reh}<T_{a, {\rm eq}}$, the axions are never in thermal equilibrium. Using eqs.~(\ref{axionprod}) and (\ref{axionthermal}), we find that the abundance of the axions produced from the inflaton scattering is suppressed as
\bea
\frac{Y_a}{Y^{\rm eq}_a}  = 0.025 \bigg(\frac{T_{\rm reh}}{T_{a, {\rm eq}}}\bigg)\bigg(\frac{10^{-11}}{\lambda_{H\Phi}}\bigg)^2\bigg(\frac{f_a}{10^{11}\,{\rm GeV}}\bigg)^{2.246}. \label{axionnonthermal}
\eea

We also remark on the effects of preheating for the axion production. As the effective mass of the axion during reheating is field-dependent as $m^2_a=\lambda_\Phi \phi^2$, as discussed in the previous subsection. Thus, the evolution of the axion becomes non-adiabatic, so the axion production from the parametric resonance can occur.  Then, the axions produced during preheating can also contribute to dark radiation. In our model, using the results in Ref.~\cite{Clery:2023ptm}, the equation governing the rescaled axion perturbation, $A_k=\omega^{1/(1-m)}a_k$, is given by
\bea
A^{\prime\prime}_k +\bigg(\kappa^2 +\frac{m^2m^2_a}{\omega^2}\bigg)A_k=0
\eea
with 
\bea
\kappa^2\equiv \frac{m^2k^2}{\omega^2 a^2}.  
\eea
Here, the prime denotes the derivative with respect to $z$ with $dz=\frac{\omega}{m}\, dt$ and the effective mass term for the axion perturbation becomes $\frac{m^2m^2_a}{\omega^2}=1.39 m^2{\cal P}^2(t)$.  Then, for $m=2$, the axion perturbation in the range, $2.78<\kappa^2<3.21$, can grow \cite{preheating}, so the axions with momentum $k\sim \omega$ can be generated by preheating, so they can give rise to dark radiation. The detailed study on the axion production during preheating is beyond the scope of our work, so we postpone the related analysis to a future work.

\section{Axion dark matter from the kinetic misalignment}

We determine the axion abundance from the axion kinetic misalignment generated during inflation and compare it with the observed relic density for dark matter. We focus on the case where the inflation is driven dominantly by the PQ conserving term.

\subsection{Evolution of the axion velocity}

From eq.~(\ref{slow-theta}), we obtain the velocity of the axion at the end of inflation, as follows,
\bea
{\dot\theta}_{\rm end}\simeq -\frac{\sqrt{2\epsilon_{\theta,{\rm end}}} H_{\rm end}}{6\sinh^2\big(\frac{\phi_{\rm end}}{\sqrt{6}M_P}\big)},
\eea
so the PQ Noether charge at the end of inflation becomes
\bea
n_{\theta,{\rm end}}=6M^2_P \sinh^2\Big(\frac{\phi_{\rm end}}{\sqrt{6}M_P}\Big) |{\dot\theta}_{\rm end}|\simeq M^2_P \sqrt{2\epsilon_{\theta,{\rm end}}} \, H_{\rm end}.
\eea
Here, $\epsilon_{\theta,{\rm end}}, H_{\rm end}, \phi_{\rm end}$ are evaluated at the end of inflation, given by 
\bea
\epsilon_{\theta,{\rm end}} = \frac{1}{2 V^2_E} \Big(\frac{\partial V_E}{\partial \theta}\Big)^2 \lesssim 1
\eea
where the PQ violating terms are taken to be smaller than the PQ conserving term. 
 
After the end of inflation, the total Noether charge for the PQ symmetry is conserved approximately, so we can take $a^3 n_\theta=a^3 \phi^2 {\dot\theta}\simeq$ const.  Then, during reheating, the inflaton condensate is radiation-like, so the radial mode scales with the scale factor by $\phi\propto a^{-1}$. Then, the axion velocity decreases by ${\dot\theta}\propto a^{-1}$. But, for a low reheating temperature for which the reheating is delayed, the quadratic term in the PQ potential becomes dominant. In this case, we need to take into account the early matter domination from the radial mode during reheating. After reheating, the inflaton settles down to the minimum of the potential, $\phi=f_a$, so the axion velocity scales by ${\dot\theta}\propto a^{-3}$. Therefore, as the kinetic energy density for the axion  is given by $\rho_\theta=\frac{1}{2}\phi^2 {\dot\theta}^2$, its scaling with the scale factor changes from $a^{-4}$ to $a^{-6}$, namely, from radiation to kination. 

As a result, the PQ Noether charge density from the axion rotation red-shifts at the end of reheating by
\bea
n_\theta(T_{\rm RH})=n_{\theta,{\rm end}}\,\bigg(\frac{a_{\rm end}}{a_{\rm RH}}\bigg)^3
\eea
where $a_{\rm end}, a_{\rm RH}$ are the values of the scale factor at the end of inflation and the reheating completion, respectively.
Then, suppose that the reheating temperature is sufficiently high such that $\phi(a_{\rm RH})>3f_a$, namely, $T_{\rm RH}>T^c_{\rm RH}$, with
\bea
T^c_{\rm RH}&\equiv& \bigg(\frac{90\lambda_\Phi}{8\pi^2 g_*}\bigg)^{1/4} 2f_a \nonumber \\
&=& \bigg(\frac{100}{g_*}\bigg)^{1/4} \bigg(\frac{f_a}{10^{11}\,{\rm GeV}}\bigg)\, (1.2\times 10^8\,{\rm GeV}).
\eea
In this case, using
\bea
\frac{a_{\rm end}}{a_{\rm RH}}=\bigg(\frac{\rho_{\rm RH}}{\rho_{\rm end}}\bigg)^{1/4}, 
\eea
with $\rho_{\rm RH}=\frac{\pi^2}{30} g_*(T_{\rm RH}) T^4_{\rm RH}$ and $\rho_{\rm end}=\frac{3}{2} V_E(\phi_{\rm end})$, we obtain
the PQ Noether charge density at the reheating temperature, as follows,
\bea
n_\theta(T_{\rm RH}) = n_{\theta,{\rm end}}\,\bigg(\frac{\pi^2 g_*(T_{\rm RH})T^4_{\rm RH}}{45 V_E(\phi_{\rm end})}\bigg)^{3/4}. \label{density1}
\eea
Here, $g_*(T_{\rm RH}), g_*(T_*)$ are the number of the effective entropy degrees of freedom at the reheating temperature and the onset of the axion oscillation, respectively. Thus, the  PQ Noether charge density at $T=T_{\rm RH}$ is independent of the reheating temperature, so is the axion abundance. 

We also remark that when the reheating is delayed such that $T_{\rm RH}<T^c_{\rm RH}$, the energy density of the inflation scales during reheating as 
\bea
\rho_\phi =\rho_{\rm end} \bigg(\frac{a_{\rm end}}{a_c}\bigg)^4 \bigg(\frac{a_c}{a_{\rm RH}}\bigg)^3
\eea
where $a_c$ is the scalar factor when $\phi(a_c)=3f_a$ such that the inflation becomes matter-like for $a>a_c$.  Then, using 
\bea
\frac{a_c}{a_{\rm RH}}=\bigg(\frac{\rho_{\rm RH}}{\rho_{\phi,c}}\bigg)^{1/3}=\bigg(\frac{T_{\rm RH}}{T^c_{\rm RH}}\bigg)^{4/3}, \label{MD}
\eea
we get the PQ Noether charge density at the end of reheating  for $T_{\rm RH}<T^c_{\rm RH}$ as 
\bea
n_\theta(T_{\rm RH})&=&n_{\theta,{\rm end}}\,\bigg(\frac{a_{\rm end}}{a_c}\bigg)^3\bigg(\frac{a_c}{a_{\rm RH}}\bigg)^3 \nonumber \\
&=&n_{\theta,{\rm end}}\ \bigg(\frac{\pi^2 g_*(T_{\rm RH}) T^4_{\rm RH}}{45 V_E(\phi_{\rm end})}\bigg)^{3/4} \bigg(\frac{T_{\rm RH}}{T^c_{\rm RH}}\bigg). \label{density2}
\eea
Here, in eqs.~(\ref{MD}) and (\ref{density2}), we took $\rho_\phi(a_{\rm RH})=\rho_{\rm RH}$ at reheating completion, $a=a_{\rm RH}$, and $\rho_{\phi,c}$ is the energy density of the inflaton at $a=a_c$, which is rewritten as $\rho_{\phi,c}=\rho_{\rm end}\big(\frac{a_{\rm end}}{a_c}\big)^4\equiv\frac{\pi^2}{30} g_*(T_{\rm RH}) (T^c_{\rm RH})^4$.
Therefore, as compared to the case when reheating is complete during the dominance of the quartic potential, the PQ Noether charge density is diluted by the extra factor, $\frac{T_{\rm RH}}{T^c_{\rm RH}}$, for a low reheating temperature. This is because the universe would have been expanded more during the early matter domination due to a smaller Hubble expansion rate. Accordingly, the PQ Noether charge density at the axion oscillation is suppressed by the low reheating temperature.

\subsection{Dark matter abundance from axions}

After the QCD phase transition, the QCD instanton effects contribute to the axion potential, so the axion is confined to one of the local minima when the kinetic energy of the axion is comparable to the potential of the axion, namely, $\frac{1}{2} f^2_a {\dot\theta}^2(T_*)=2m^2_a(T_*) f^2_a $ at the temperature $T=T_*$.
Thus, we need ${\dot\theta}(T_*)=2m_a(T_*)$ and $m_a(T_*) \geq 3 H(T_*)$ for the axion oscillation \cite{Co:2019jts}. Therefore, we obtain the condition for  the axion kinetic misalignment as ${\dot\theta}(T_*)\geq 6H(T_{\rm osc})$. Here, we remark that the axion kinetic misalignment delays the onset of oscillation, namely,  $T_*\leq T_{\rm osc}$. Here, $T_{\rm osc}$ is the temperature of the standard axion oscillation determined by $m_a(T_{\rm osc})=3H(T_{\rm osc})$, which is given by
\bea
T_{\rm osc} =\bigg(\frac{10}{\pi^2 g_*}\bigg)^{1/12}\Big(m_a(0) M_P \Lambda^4_{\rm QCD}\Big)^{1/6}
\eea
where $m_a(0)$ is the axion mass at zero temperature whose precise value is given \cite{axionmass} by
\bea
m_a(0)= 5.691(51)\times 10^{-3}\,{\rm eV} \bigg(\frac{10^9\,{\rm GeV}}{f_a}\bigg).
\eea
For instance, for $f_a=10^9\,{\rm GeV}$, $\Lambda_{\rm QCD}=150\,{\rm MeV}$ and $g_*=75.75$, we determine the temperature at the axion oscillation as $T_{\rm osc}=3.05\,{\rm GeV}$.

For the axion kinetic misalignment to be a dominant mechanism for determining the axion relic density, we obtain the axion relic abundance by
\bea
\Omega_a h^2 = 0.12 \bigg(\frac{10^9\,{\rm GeV}}{f_a}\bigg) \bigg(\frac{Y_\theta}{40}\bigg)  \label{relic}
\eea
where $Y_\theta$ is the abundance for the axion  given by $Y_\theta=\frac{n_\theta(T_{\rm RH})}{s(T_{\rm RH})}$ with $n_\theta(T_{\rm RH})$  and  $s(T_{\rm RH})$ being the Noether charge density and the entropy density at reheating, respectively.  Here, we used $\frac{\rho_a}{s}=Cm_a(0) Y_\theta$ with $C\simeq 2$ to convert the PQ charge abundance to the relic density \cite{Co:2019jts}.Here, we note that there is an uncertainty of order one in  $C$ due to the particle production from the axion fragmentation \cite{axionfrag}. So, we would need to rely on the lattice simulations for a more precise calculation of the axion kinetic misalignment, which is beyond the scope of our work.

In the usual misalignment mechanism, we note that the axion abundance is determined by $Y_{a,{\rm mis}}=\frac{n_a}{s}$ at the onset of the axion oscillation at $T=T_{\rm osc}$ \cite{Bae:2008ue,Wantz:2009it,GrillidiCortona:2015jxo,Borsanyi:2016ksw}, namely, 
\bea
Y_{a,{\rm mis}}=0.11 \bigg(\frac{f_a}{10^9\,{\rm GeV}}\bigg)^{13/6}.
\eea
Then, the dominance with the axion kinetic misalignment, namely, $Y_\theta>Y_{a,{\rm mis}}$,  requires $f_a<1.5\times 10^{11}\,{\rm GeV}$ \cite{Co:2019jts}.

\begin{figure}[!t]
\begin{center}
 \includegraphics[width=0.45\textwidth,clip]{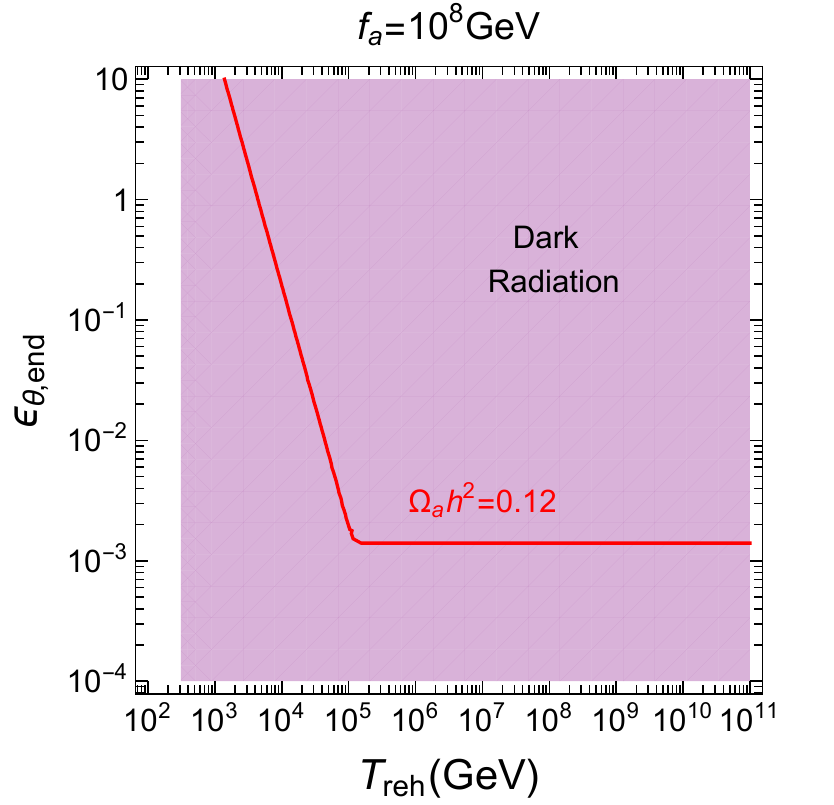}\,\,  \includegraphics[width=0.45\textwidth,clip]{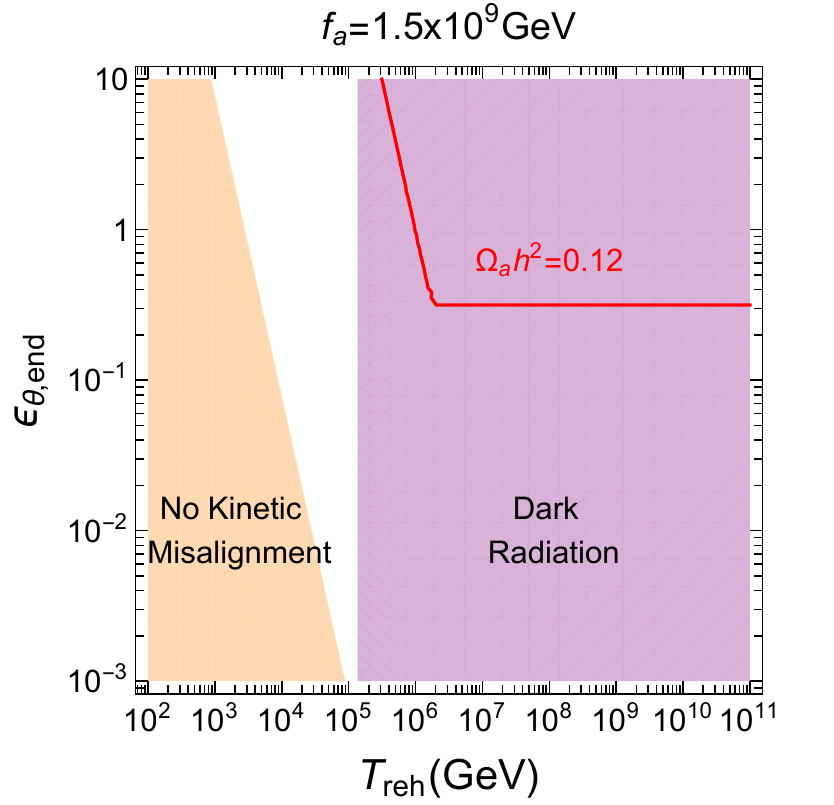}  
 \end{center}
\caption{Parameter space in the reheating temperature $T_{\rm reh}$ vs $\epsilon_{\theta,{\rm end}}$ for axion dark matter with kinetic misalignment. The correct relic density is obtained by the axion kinetic misalignment along the red line. The kinetic misalignment is sub-dominant in orange region, and the axions produced from the inflaton scattering were in thermal equilibrium, becoming dark radiation at a detectable level in purple regions. We took $f_a=10^8\,{\rm GeV}, 1,5\times 10^{9}\,{\rm GeV}$ on the left and right plots, respectively. }
\label{fig:relic1}
\end{figure}

In Fig.~\ref{fig:relic1}, we depict the parameter space for the reheating temperature and the slow-roll parameter for the axion, $\epsilon_{\theta,{\rm end}}$, satisfying the correct relic density by the axion kinetic misalignment in red lines. We show that the axions produced from the saxion scattering  becomes thermalized after reheating and provides a detectable dark radiation in purple regions, whereas the kinetic misalignment is sub-dominant in the orange region. The correct relic density is achievable even for a relatively small axion decay constant such as $f_a=10^8\,{\rm GeV}$ and $1.5\times 10^{9}\,{\rm GeV}$ in the left and right plots, respectively. Here, for the axion relic density, we used eq.~(\ref{relic}) with eqs.~(\ref{density1}) or (\ref{density2}), depending on whether $T_{\rm RH}>T^c_{\rm RH}$ or not, and took $V_E(\phi_{\rm end})=0.089 V_I$ with the inflaton potential energy $V_I$ being constrained by the CMB normalization in eq.~(\ref{CMB}) and $\phi_{\rm end}=1.5M_P$. 

It is worthwhile to remark that the reheating temperature corresponding to the correct relic density is achieved from the inflaton decay with the Yukawa coupling $y_f$ in eq.~(\ref{decay}), and/or  the inflaton scattering with the Higgs-portal coupling $\lambda_{H\Phi}$ in eq.~(\ref{scattering}).  A low reheating temperature below $T_{\rm RH}\sim 10^4\,{\rm GeV}$ can be obtained from the inflaton decay with $y_f\lesssim 10^{-4}$, but a high reheating temperature  up to $T_{\rm RH}\sim 10^{11}\,{\rm GeV}$ is possible due to the inflaton scattering with $\lambda_{H\Phi}\lesssim 10^{-7}$, being consistent with a small running quartic coupling for the PQ field. We also note that eq.~(\ref{Neff}) is sufficient for computing the dark radiation abundance in the parameter space explaining the correct relic density. This is because axions are thermalized with the SM plasma, namely, $T_{\rm RH}\gtrsim 310 \,{\rm GeV}, 1.4\times 10^5\,{\rm GeV}$ is satisfied for $f_a=10^8\,{\rm GeV}, 1.5\times 10^9\,{\rm GeV}$, respectively, and the dark radiation becomes independent of the initial abundance produced from the inflaton scattering given in eq.~(\ref{axionnonthermal}).

\begin{figure}[!t]
\begin{center}
 \includegraphics[width=0.45\textwidth,clip]{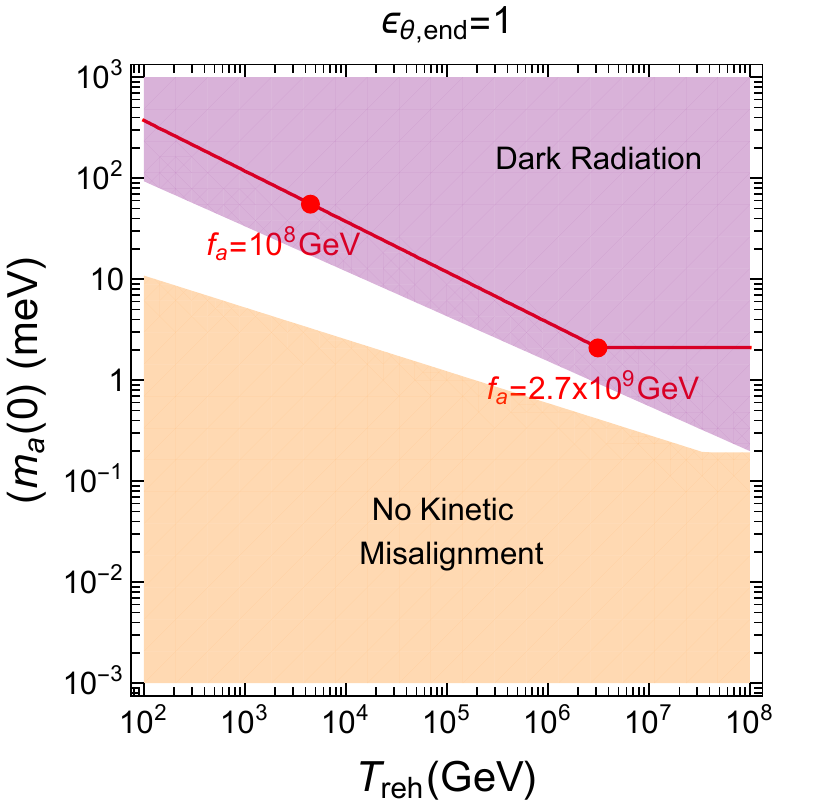}\,\,  \includegraphics[width=0.45\textwidth,clip]{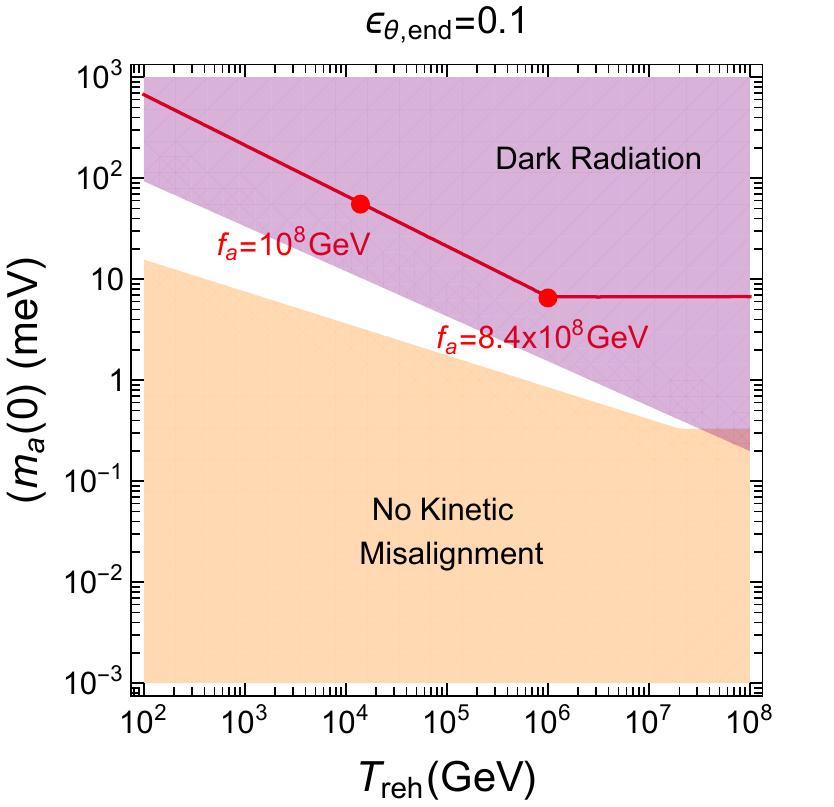}  
 \end{center}
\caption{Parameter space in the reheating temperature $T_{\rm reh}$ vs the axion mass $m_a(0)$  for axion dark matter with kinetic misalignment. The correct relic density is obtained by the axion kinetic misalignment along the red line. The kinetic misalignment is sub-dominant in orange regions and the axions produced from the inflaton scattering were in thermal equilibrium, becoming dark radiation at a detectable level in purple regions. We took $\epsilon_{\theta,{\rm end}}=1, 0.1$ on the left and right plots, respectively. }
\label{fig:relic2}
\end{figure}

In Fig.~\ref{fig:relic2}, we show the correlation between the reheating temperature and the axion mass at zero temperature, accounting for the correct relic density from the axion kinetic misalignment in red lines as in Fig.~\ref{fig:relic1}. We chose the slow-roll parameter of the axion at the end of inflation to $\epsilon_{\theta,{\rm end}}=1, 0.1$ in the left and right plots, respectively. We show the corresponding axion decay constant with $f_a\geq 10^8\,{\rm GeV}$ satisfying the correct relic density for dark matter and the reheating temperature needs to be above $T_{\rm reh}=4\times10^3\,{\rm GeV}$ for $\epsilon_{\theta,{\rm end}}=1$, and above $T_{\rm reh}=10^4\,{\rm GeV}$  for $\epsilon_{\theta,{\rm end}}=0.1$, respectively. Here, the lower end of $f_a$ is taken from the bounds from Supernova 1987A \cite{SN,SN2}, which is $f_a>(1-4)\times 10^8\,{\rm GeV}$. As in Fig.~\ref{fig:relic1}, the region with a detectable dark radiation  is shown in purple regions and  the kinetic misalignment is sub-dominant in orange regions.

\section{Conclusions}

We have presented a consistent model of the PQ inflation for realizing the kinetic misalignment of the QCD axion. The radial component of the PQ field drives inflation near the pole of the kinetic term in the Einstein frame, leading to a successful prediction for inflationary observables with a small quartic coupling for the PQ field during inflation. The angular component of the PQ field, namely, the QCD axion, receives a nonzero initial velocity during inflation, due to the PQ violating potential. 

We performed the analysis of the perturbative reheating in the presence of the couplings of the PQ field to the SM Higgs as well as the extra vector-like quark responsible for PQ anomalies. We found that a sufficiently large reheating temperature is achieved from the decays and scattering processes of the inflaton and the axions produced from the inflation scattering can become dark radiation at a detectable level in the future CMB experiments if they are thermalized. Thus, as the inflaton reaches the minimum of the potential at reheating completion, the approximately conserved Noether charge for the PQ symmetry survives until the time of the axion oscillation, making the axion kinetic misalignment to be a dominant mechanism for axion dark matter. 

Taking the PQ violating potential to be sub-dominant during inflation and free from the axion quality problem in the vacuum, we set the maximum value of the axion velocity at the end of inflation and found the consistent parameter space for the axion kinetic misalignment in the reheating temperature, the initial velocity of the axion and the axion decay constant. For the fixed initial velocity of the axion, the larger the axion decay constant, the smaller the maximum reheating temperature for the axion kinetic misalignment. To be consistent with the astrophysical bounds on the axion couplings and the axion kinetic misalignment, we took the axion decay constant above $f_a=10^8\,{\rm GeV}$ to satisfy the bounds from Supernova 1987A,and showed that the maximum temperature needs to be above $T_{\rm reh}=4\times 10^3\,{\rm GeV}-10^4\,{\rm GeV}$, for the slow-roll parameter for the axion at the end inflation, $\epsilon_{\theta,{\rm end}}=0.1-1$.

\section*{Acknowledgments}

We are supported in part by Basic Science Research Program through the National
Research Foundation of Korea (NRF) funded by the Ministry of Education, Science and
Technology (NRF-2022R1A2C2003567 and NRF-2021R1A4A2001897). The work of MJS was also supported by the Chung-Ang University Graduate Research Graduate Research Scholarship in 2023. AGM is supported in part by the CERN-CKC PhD student fellowship program and HML and AGM acknowledge support by CERN Theory department during their stay and support by Institut Pascal at Universit\'e Paris-Saclay during the Paris-Saclay Astroparticle Symposium 2023.

%\def\theequation{A.\arabic{equation}}

%\setcounter{equation}{0}

%\vskip0.8cm
%\noindent
%{\Large \bf Appendix A: The general vacuum for waterfall fields}

\end{document}